# *In-Vivo* Hyperspectral Human Brain Image Database for Brain Cancer Detection

**HIMAR FABELO**[1], **SAMUEL ORTEGA**[1], **ADAM SZOLNA**[2], **DIEDERIK BULTERS**[3], **JUAN F. PIÑEIRO**[2], **SILVESTER KABWAMA**[3], **ARUMA J-O'SHANAHAN**[2], **HARRY BULSTRODE**[4], **SARA BISSHOPP**[2], **B. RAVI KIRAN**[5], **DANIELE RAVI**[6], **RAQUEL LAZCANO**[7], **DANIEL MADROÑAL**[7], **CORALIA SOSA**[2], **CARLOS ESPINO**[2], **MARIANO MARQUEZ**[2], **MARÍA DE LA LUZ PLAZA**[8], **RAFAEL CAMACHO**[8], **DAVID CARRERA**[2], **MARÍA HERNÁNDEZ**[2], **GUSTAVO M. CALLICÓ**[1], **(Member, IEEE), JESÚS MORERA MOLINA**[2], **BOGDAN STANCIULESCU**[9], **GUANG-ZHONG YANG**[6], **(Fellow, IEEE), RUBÉN SALVADOR**[7], **EDUARDO JUÁREZ**[7], **(Member, IEEE), CÉSAR SANZ**[7], **(Senior Member, IEEE), AND ROBERTO SARMIENTO**[1]

[1]Institute for Applied Microelectronics, University of Las Palmas de Gran Canaria, 35017 Las Palmas de Gran Canaria, Spain
[2]Department of Neurosurgery, University Hospital Doctor Negrin, 35010 Las Palmas de Gran Canaria, Spain
[3]Wessex Neurological Centre, University Hospital Southampton, Southampton SO16 6YD, U.K.
[4]Department of Neurosurgery, Addenbrookes Hospital, University of Cambridge, Cambridge CB2 0QQ, U.K.
[5]Autonomous Systems, AKKA Technologies, 75008 Paris, France
[6]The Hamlyn Centre, Imperial College London, London SW7 2AZ, U.K.
[7]Centre of Software Technologies and Multimedia Systems, Universidad Politécnica de Madrid, 28031 Madrid, Spain
[8]Department of Pathological Anatomy, University Hospital Doctor Negrin, 35010 Las Palmas de Gran Canaria, Spain
[9]Ecole Nationale Supérieure des Mines de Paris, MINES ParisTech, 75006 Paris, France

Corresponding author: Himar Fabelo (hfabelo@iuma.ulpgc.es)

This work was supported in part by the European Project HELICoiD ''HypErspectraL Imaging Cancer Detection'' under Grant 618080, funded by the European Commission through the FP7 FET (Future and Emerging Technologies) Open Programme ICT-2011.9.2, in part by the ITHaCA Project ''Hyperespectral Identification of Brain Tumors'' under Grant ProID2017010164, funded by the Canary Islands Government through the Canarian Agency for Research, Innovation and the Information Society (ACIISI), in part by the 2016 Ph.D. Training Program for Research Staff of the University of Las Palmas de Gran Canaria. The work of S. Ortega was supported in part by the pre-doctoral grant given by the ''Agencia Canaria de Investigacion, Innovacion y Sociedad de la Información (ACIISI)'' of the ''Conserjería de Economía, Industria, Comercio y Conocimiento'' of the ''Gobierno de Canarias'', which is part-financed by the European Social Fund (FSE) (POC 2014-2020, Eje 3 Tema Prioritario 74 (85%)).

**ABSTRACT** The use of hyperspectral imaging for medical applications is becoming more common in recent years. One of the main obstacles that researchers find when developing hyperspectral algorithms for medical applications is the lack of specific, publicly available, and hyperspectral medical data. The work described in this paper was developed within the framework of the European project HELICoiD (*HypErspectraL Imaging Cancer Detection*), which had as a main goal the application of hyperspectral imaging to the delineation of brain tumors in real-time during neurosurgical operations. In this paper, the methodology followed to generate the first hyperspectral database of *in-vivo* human brain tissues is presented. Data was acquired employing a customized hyperspectral acquisition system capable of capturing information in the Visual and Near InfraRed (VNIR) range from 400 to 1000 nm. Repeatability was assessed for the cases where two images of the same scene were captured consecutively. The analysis reveals that the system works more efficiently in the spectral range between 450 and 900 nm. A total of 36 hyperspectral images from 22 different patients were obtained. From these data, more than 300 000 spectral signatures were labeled employing a semi-automatic methodology based on the spectral angle mapper algorithm. Four different classes were defined: normal tissue, tumor tissue, blood vessel, and background elements. All the hyperspectral data has been made available in a public repository.

The associate editor coordinating the review of this manuscript and approving it for publication was Bora Onat.

## I. INTRODUCTION

Hyperspectral imaging (HSI) is a non-contact, non-ionizing and non-invasive sensing technique suitable for medical applications [1]. HSI consists in collecting and processing information across the electromagnetic spectrum, beyond the human eye capabilities that cover the range from 390 to 700 nm [2]. HSI increases the amount of information acquired from a certain scene, compared with a conventional RGB (Red, Green and Blue) image or a multi-spectral image, by capturing more data in a large number of contiguous and narrow spectral bands over a wide spectral range. Depending on the type of sensor employed, hyperspectral (HS) cameras will cover different spectral ranges. CCD (Charge-Coupled Device) sensors cover the VNIR (Visible and Near-InfraRed) range from 400 to 1000 nm, while InGaAs (Indium Gallium Arsenide) sensors are able to capture HS images in the NIR (Near-InfraRed) range, between 900 and 1700 nm. Other types of sensors can reach larger spectral ranges. For example, MCT (Mercury Cadmium Telluride) sensors are able to acquire HS images in the SWIR (Short-Wavelength InfraRed) range, from 1000 to 2500 nm [3].

HSI has been employed for many years in remote sensing applications [4]. However, more recently, HSI is progressively being used in other fields, such as drug analysis [5], [6], food quality inspection [7]–[10] or defense and security [11], [12] among many others. Furthermore, HSI is an emerging image modality in the medical field, offering promising results for the detection and diagnosis of different types of diseases [13]. In particular, there are several studies considering cancer analysis and diagnosis. As cancer involves a change in cellular pathology associated with different chemical signature [14], it should be detected as a change in the HS signature of the tissue. HSI has been used to detect different types of cancer, such us cervix [15], breast [16]–[18], colon [19]–[22], gastrointestinal [23], skin [24], [25], prostate [26], [27], oral tissue [28] and tongue [29]–[31] cancers. Regarding HSI to assist brain surgery, the work performed by this research group in [32] is the first study reported to the best of authors' knowledge. One of the major benefits of this technology is that it can be used as a guidance tool during brain tumor resections. Unlike other tumors, brain tumor infiltrates the surrounding normal brain tissue and thus their borders are indistinct and extremely difficult to identify to the surgeon's naked eye. The surrounding normal brain tissue is critical and there is no redundancy, as in many other organs where the tumor is commonly resected together with an ample surrounding block of normal tissue. This is not possible in the brain, where it is essential to accurately identify the margins of the tumor to resect as less healthy tissue as possible [33]. As a result, HSI can be applied for a precise localization of malignant tumors during surgical procedures [32], [34].

Currently, several imaging modalities are employed as guidance tools to assist the tumor resection in neurosurgical operations. However, these systems have many limitations. Intraoperative neuronavigation systems uses preoperative image information, such as magnetic resonance imaging (MRI) or computed tomography (CT), to monitor the surgery in real-time. Nevertheless, the tumor boundary delimitation is not accurate enough due to the brain shift phenomenon [35]–[38]. On the contrary, intraoperative MRI solves the brain shift problem accurately monitoring the tissue resection intraoperatively. However, this method requires a specific operating theater with MRI-compatible equipment and increase the duration of the surgery [39]. Moreover, the refreshing rate of intraoperative MRI is much lower than the refreshing rate of HSI. On the other hand, ultrasound (US) is inexpensive, real-time, unaffected by brain shift, and for most gliomas it can identify the contrast-enhancing portion seen on MRI, reliably identifying tumor margins [40], [41]. However, it has been reported that the use of intraoperative US can cause the resection of histologically-normal parenchyma [42]. Furthermore, intraoperative US leaves borders difficult to demarcate [43] and it is also time-consuming and operator-dependent [41], requiring large experience by the user to interpret the ultrasound images. It also lacks image resolution [44]. Finally, the main competitors of HSI for brain tumor detection are the systems based on fluorescent tumor markers, such as 5-aminolevulinic acid (5-ALA). These systems are capable of providing intraoperative tumor margins detection with high accuracy and refreshing rate. However, this technique may produce important side effects for the patient, being not recommended for pediatric cases [45]. Furthermore, it can only be used to detect high-grade tumors [46]. For these reasons, HSI could lead to a potential solution to these problems, being a non-contact and label-free technique totally harmless for the patient.

This paper presents the first in-vivo HS human brain image database, mainly generated to study its ability to delineate and identify brain tumors during surgical operations using this non-invasive and non-ionizing image modality. This database has been generated within the HELICoiD (HypErspectraL Imaging Cancer Detection) project. HELICoiD was an European collaborative project funded by the Research Executive Agency (REA), through the Future and Emerging Technologies (FET-Open) Programme, under the 7th Framework Programme of the European Union. The project was a collaboration between four universities, three industrial partners and two hospitals, where the main goal was to employ HSI to develop a methodology to discriminate between tumor and normal brain tissue in surgical-time during surgical procedures by employing machine learning techniques [47]. The integration of HSI in an intraoperative image guided surgery system could have a direct impact on the patient outcomes. Potential benefits would include allowing confirmation of complete resection during the surgical procedures, avoiding complications due to the brain shift phenomenon, and providing confidence that the goals of the surgery have been successfully achieved.

## II. THE INTRA-OPERATIVE HS ACQUISITION SYSTEM

A customized intraoperative HS acquisition system [32], [48], [49], developed within the HELICoiD research project, was used to generate the in-vivo HS human brain image database. Fig. 1.a shows the intraoperative HS acquisition system being used during a neurosurgical operation at the University Hospital Doctor Negrin of Las Palmas de Gran Canaria, Spain. A Hyperspec® VNIR A-Series camera (HeadWall Photonics Inc., Fitchburg, MA, USA) composes the acquisition system. The HS camera is based on pushbroom technique and covers the spectral range from 400 to 1000 nm, with a spectral resolution of 2–3 nm, being able to capture 826 spectral bands and 1004 spatial pixels per line (Fig. 1.b and c). The camera uses a silicon CCD detector array with a minimum frame rate of 90 fps (frames per second). In this context, a frame refers to a line of 1004 pixels and 826 spectral bands. A Xenoplan 1.4 lens (Schneider Optics, Hauppauge, NY, USA) with 22.5 mm of focal length and a broadband coating for the spectral range between 400 and 1000 nm is coupled to the camera.

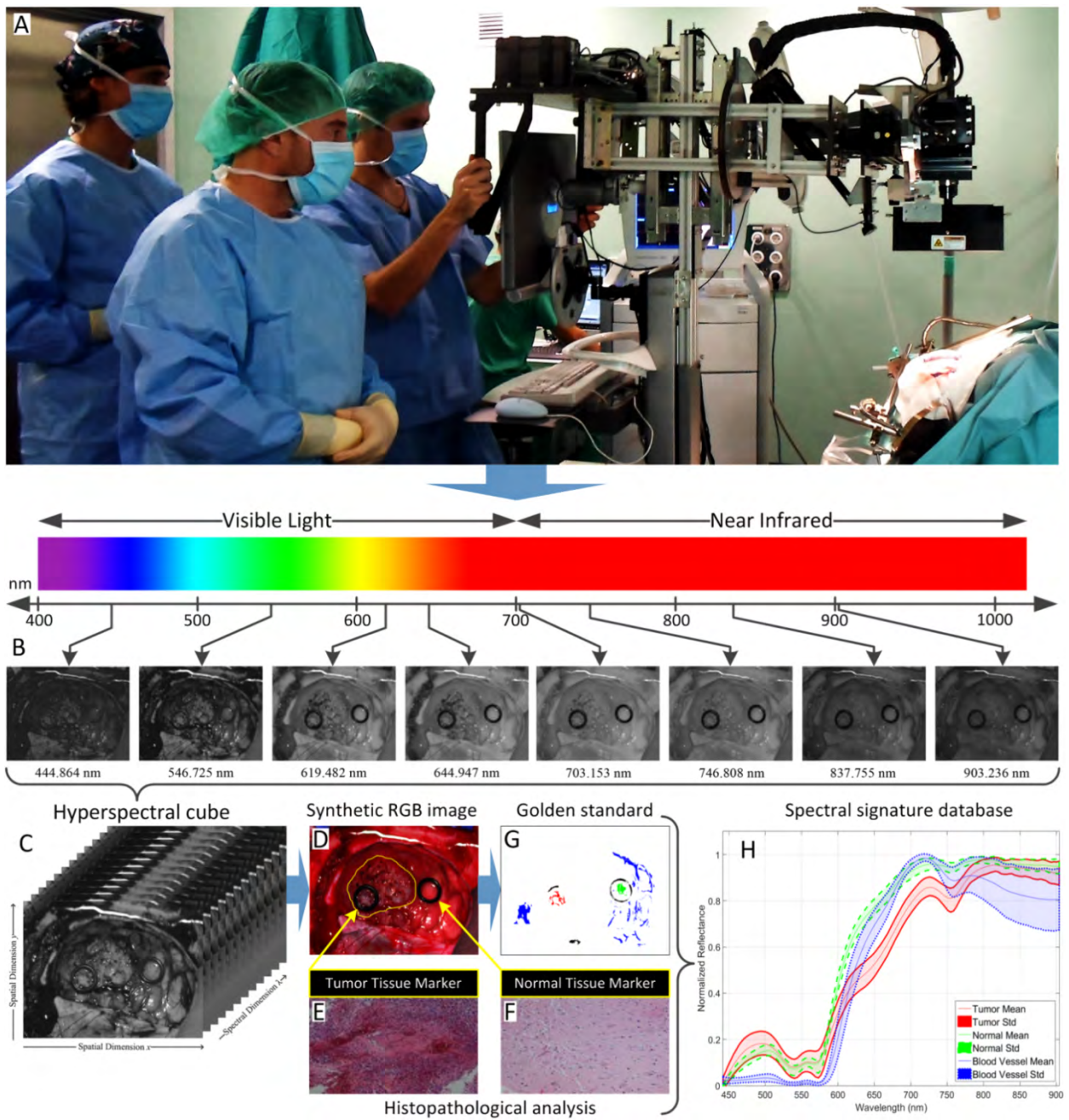

**FIGURE 1.** *In-vivo* HS brain surface acquisition procedure. (a) Hyperspectral acquisition system being used during the acquisition process in a neurosurgical operation. (b) Hyperspectral images acquired with the acquisition system at different wavelengths from a patient affected by a glioblastoma tumor. (c) HSI data cube. (d) RGB image generated from the HS cube with the tumor tissue marker (left) and the normal tissue marker (right) placed on the brain surface. (e) and (f) Histopathological images of the tumor tissue sample (glioblastoma) and normal tissue sample respectively. (g) Gold standard map where certain pixels have been labeled in four different classes: normal brain tissue (green), tumor tissue (red), blood vessel (blue) and background (black). (h) Average and standard deviation (Std) of the pre-processed spectral signatures of tumor tissue, normal tissue and blood vessel labeled pixels, represented in red, green and blue color respectively.

The camera is fixed to a scanning platform composed by a stepper motor coupled to a spindle capable of covering an effective capturing area of 230 mm. The scanning platform is required to obtain the second spatial dimension of the HS cube, since the pushbroom camera only captures a single spatial dimension. Additionally, the acquisition system integrates an illumination system based on a QTH (Quartz-Tungsten-Halogen) lamp connected to a cold light emitter through fiber optic cable that cover the spectral range between 400 and 2200 nm. This illumination system allows providing cold illumination over the brain surface in order to avoid the high temperature produced by the QTH lamp.

## III. *IN-VIVO* HS HUMAN BRAIN DATABASE
### A. PARTICIPANTS

All adult patients (18 years old and above) undergoing craniotomy for resection of intra-axial brain tumors at both participating sites were approached for inclusion in the study. The data acquisition was performed in two different campaigns that covered the period comprised from March 2015 to June 2016. The University Hospital of Southampton (UHS), UK, and the University Hospital Doctor Negrin (UHDRN) of Las Palmas de Gran Canaria, Spain, were the participant sites. Patients with both primary and secondary tumors were included in the database. Patients that underwent resection of meningioma, where the dura was resected, were also included whenever it was possible to properly capture the exposed normal brain. The study protocol and consent procedures were approved by the *Comité Ético de Investigación Clínica-Comité de Ética en la Investigación* (CEIC/CEI) for the University Hospital Doctor Negrin and the National Research Ethics Service (NRES) Committee South Central - Oxford C for the University Hospital of Southampton. Written informed consent was obtained from all the subjects.

### B. HS IMAGE ACQUISITION PROCEDURE

This section provides an overview of the procedure carried out to obtain the HS *in-vivo* images from the human brain surface that were stored in the database (Fig. 1). Furthermore, the process of labeling the samples as tumor tissue or normal brain tissue is described. The procedure followed to acquire and label the spectral signatures of the *in-vivo* brain surface is based on five steps: patient preparation, HS image acquisition, tissue resection, neuropathology evaluation and sample labeling.

#### 1) PATIENT PREPARATION

Before the operation, a CT (Computed Tomography) and MRI (Magnetic Resonance Imaging) of the patient's head compatible with IGS (Image Guide Stereotactic) are performed. These images are uploaded onto the IGS system before the operation started. Then, the patient is placed in a supine position, under general anesthesia and is registered to the IGS system. A scalp incision is made and a burr hole/s drill using a high-speed drill. A craniotome is then inserted into the burr hole/s and a bone flap cut out (craniotomy). The dura is then cut with a knife (durotomy) to expose the brain surface.

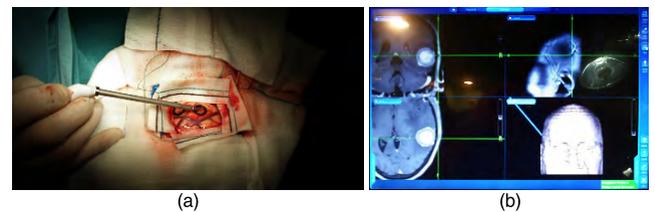

**FIGURE 2.** MRI markers registration. (a) IGS system pointer over the tumor marker located on the exposed brain surface. (b) IGS system screen snapshot with the coordinates of the tumor marker in the MRI.

#### 2) HS IMAGE ACQUISITION

HS images are captured after durotomy, before the arachnoid and pia are breached in cases where the tumor extended to the brain surface. Sterilized rubber ring markers were placed to identify the position of the tumor and the normal brain (Fig. 1.d), employing the IGS system pointer to establish the exact location of the markers over the brain. This pointer provides information about the position of the markers with respect to an MRI or CT, which are performed prior to the surgical procedure.

Fig. 2.a and b show the use of the IGS system pointer to identify the position of the markers over the brain on MRI. Normal brain markers are also placed by the operating surgeon based on visual appearance and anatomical relationship to sulci and gyri, apart from the IGS feedback. Normal markers are placed where the operating surgeon is quite confident that the brain tissue is normal. The operator of the HS acquisition system then captures the HS image of the exposed brain surface. The image size established by the operator for the image acquisition should be oversized to avoid the inherent movement of the system when the scanning platform is starting the capturing process. This will ensure the spatial coherence of the center of the image where the region of interest (the exposed brain surface) is located. In addition, the operator must avoid unintentional movements of the acquisition system during image acquisition for the same reason. In case of an unexpected movement, the capturing process must be repeated.

#### 3) TISSUE RESECTION

After the first HS image capture, the HS acquisition system is moved out from the surgical area and the operating surgeons undertake the tumor resection. Surgeons take samples of the tissue located inside the tumor marker and place it in a sterile container. An identification (ID) number is assigned to the container with the corresponding marker number. These samples are sent to the neuropathology laboratory and the results are used for the gold standard dataset generation. The IDs assigned to the markers serve as guidance in the labeling process.

When possible, a second set of images is captured while the tumor is being resected. When there is macroscopically normal brain and tumor exposed and when the operating surgeon feels that it is safe to temporarily hold surgery, the surgeon ensures perfect hemostasis and washes the field thoroughly with warm saline to wash away any residual blood while ensuring no significant temperature change (and resultant blood flow change). Next, the field is sucked dry by the application of a cottonoid to the parenchyma and applying suction to this. Then, the operating surgeon identifies the most suitable location to capture the images.

### 4) NEUROPATHOLOGY EVALUATION

The resected tissue is sent to the neuropathology laboratory where it is formalin fixed and undergoes standard H&E (Hematoxylin and Eosin) staining and any further required staining to establish a definitive histopathological diagnosis. Neuropathologists perform the histopathological diagnosis, employing techniques used in the routine clinical practice. Samples are classified as tumor or normal brain. Furthermore, tumor samples are subdivided into tumor type and grade. Fig. 1.e and f show an example of a histopathological image of a glioblastoma and normal brain tissues, respectively.

### 5) SAMPLE LABELING

In the last step, by using the information provided by the neuropathologists and the knowledge of the operating surgeons, several pixels are labeled taking as a reference the spectral signatures of certain pixels inside the markers and outside the markers for the non-tumor classes. The total number of reference pixels selected per image is detailed in Table 1. After the labeling process (explained in section IV), a gold standard map of each image is obtained (Fig. 1.g). Pixels that present spectral reflections produced by the non-uniformity of the brain surface are avoided.

The gold standard maps are composed of four classes: tumor tissue (red color), normal brain tissue (green color), blood vessels (blue color) and background (black color). Based on the indications given by the pathologist, the eventually possible normal inflamed labeled tissue has been included in the normal class. The background class comprises any other tissues, materials or substances that can be present in the surgical scene and are not relevant for the tumor resection. Blood vessels and background pixels are labeled by the surgeon with the naked eye. White pixels in the gold standard map are the pixels that are not labelled.

### C. DATA PRE-PROCESSING

The HS data obtained are pre-processed following the pre-processing chain presented in [32]. Four main steps compose this chain: image calibration, noise filtering, band averaging and pixel normalization.

After image acquisition, the HS raw data are calibrated using white and dark reference images. The white reference image is captured from a Spectralon® tile (SphereOptics GmbH, Herrsching, Germany) in the same illumination conditions in which the images were captured. This material reflects the 99% of the incoming radiation in the range between 400 and 1000 nm. The dark reference image is obtained by keeping the camera shutter closed. This calibration procedure is performed to avoid the problems of the spectral non-uniformity of the illumination device and the dark currents produced by the camera sensor, avoiding the influence of the external illumination conditions. Next, due to the high noise generated by the camera sensor, a set of steps with the goal of removing this noise from the spectral signatures and to reduce the number of bands of the samples without losing relevant spectral information are applied.

Fig. 1.h shows an example of the average and standard deviations of the pre-processed spectral signatures of a glioblastoma and normal brain tissues as well as blood vessel labeled pixels, represented in red, green and blue respectively.

## IV. HS GOLD STANDARD DATABASE

In this application where human living patients are involved, it is not possible to achieve a complete gold reference map of the captured image with 100% of certainty that the pixel represents the established class. To achieve that, a pathologist should analyze the entire brain tissue exposed in the image and this is obviously not possible due to ethical reasons, since in this case the neurosurgeon should resect all the tissue exposed in the brain surface (including tumor and normal), causing serious problems to the patient health. In other fields such as remote sensing or even in the medical field but using ex-vivo or in-vitro tissue the complete gold reference generation is a relatively easy task, but using in-vivo human samples (and especially in the brain) this task is highly complex and nearly impossible nowadays.

In this section, we propose a methodology to achieve a partial gold standard map of the HS image of the in-vivo brain surface based on four key factors: 1) The pathological analysis of a tumor sample to confirm the tumor tissue location and diagnosis; 2) The association of the HS captured image with the intraoperative neuronavigation system to locate the approximated area of the tumor; 3) The experience and knowledge of the neurosurgeon to identify the normal, hypervascularized and background classes and also the approximated area of the tumor tissue following the indications of the neuronavigation system and the pathological diagnosis; 4) The spectral properties of each tissue class to find the most similar pixels with respect to a reference pixel selected by the operating surgeon by using a labelling tool specifically developed to this end.

Commonly, manual gold standard dataset definition is usually done by visual inspection of the scene and the successive labeling of each sample. This labeling methodology can produce errors in the final gold standard map (especially in this application), since it is possible to label by mistake pixels with different spectral information in the same class due to the manual procedure. Therefore, in order to build a gold standard as reliable as possible, a labelling tool based on the Spectral Angle Mapper (SAM) algorithm [50] was developed for this application to generate a robust and efficient gold standard dataset.

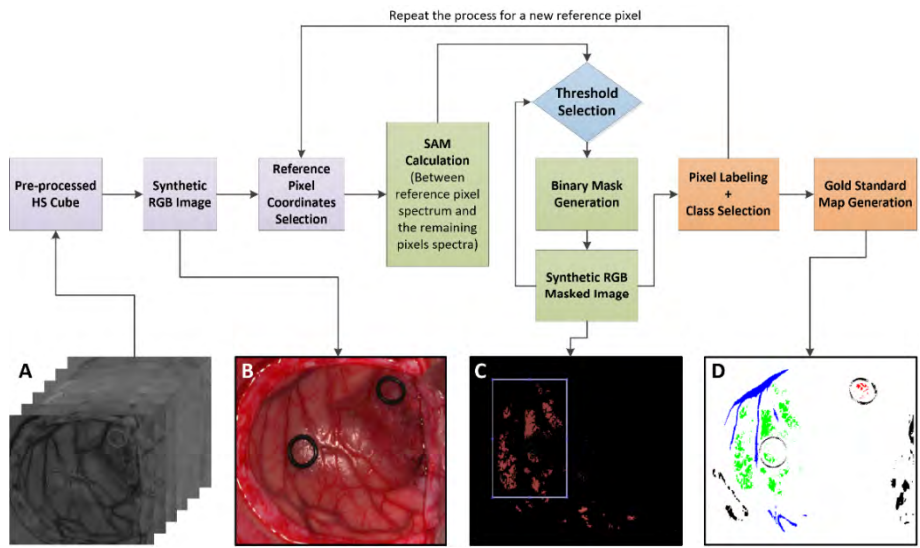

**FIGURE 3.** Semi-automatic labeling process flowchart. (a) Pre-Processed HS cube. (b) Synthetic RGB image extracted from the HS cube. (c) SAM mask over the synthetic RGB image. (d) Final gold standard map.

An interactive graphical user interface was designed using the Matlab® GUIDE application (The MathWorks Inc., Natick, MA, USA). In the first step of the labeling process, the specialist (the operating surgeon) selects the coordinates of a reference pixel in a synthetic RGB image generated from the HS cube. The synthetic RGB image is generated by extracting three specific spectral bands from the HS cube that correspond with red (708.97 nm), green (539.44 nm), and blue (479.06 nm) colors. The reference pixel can be selected at the site where the biopsy was performed (where the marker is located) or at a site far enough from the tumor margins where the surgeon can be quite confident that the tissue is abnormal (in the case of tumor labeling). The tissue presented inside the tumor markers was sent to pathology for a precise diagnosis of the tumor. In the case of normal tissue, hypervascularized tissue and background classes, the labeling is performed by selecting a reference pixel by the naked eye based on the surgeon's knowledge and experience. Then, the spectral angle between the selected pixel and the other pixels of the HS cube is calculated. In this moment, the operating surgeon manually establishes a threshold to match the physiological features of the selected tissue. This means that the most spectrally similar pixels with respect to the selected reference pixel are highlighted using a binary mask. This masked image contains only pixels with a spectral angle lower than the established threshold with respect to the reference pixel. Once the user concludes that only the pixels belonging to one class have been highlighted, the selected pixels are assigned to that class. Neurosurgeons were instructed to select only few sets of very reliable pixels instead of a wider set of uncertain pixels.

This labeling framework provides two main advantages to generate the gold standard maps. On the one hand, when the specialist selects a reference pixel, it is possible to ensure that the selected pixel indeed belongs to a certain class by looking at the synthetic RGB masked image, where the pixels with lower spectral angle with respect to the reference pixel are shown. On the other hand, the process of manually selecting pixels from a HS cube for each class is a time-consuming task, so this semi-automatic method allows generating the gold standard in an efficient way.

Fig. 3 presents the block diagram of this procedure and the information available at each stage of the labeling process. Fig. 3.a represents the pre-processed HS cube that is the input to the labeling chain while Fig. 3.b shows the synthetic RGB representation of the HS cube, where the reference pixels are selected. Fig. 3.c illustrates the synthetic RGB masked image after applying the SAM algorithm between the reference pixel and the other pixels of the HS cube with a certain fixed threshold. In addition, in this image is possible to see the area selected by the specialist, which will be labeled in a certain class. Finally, Fig. 3.d shows the final gold standard map generated after the labeling procedure, where the labeled pixels that belong to tumor tissue, normal brain tissue, blood vessels and background are identified with red, green, blue and black colors respectively. Each HS image of the generated database has its respective gold standard map. In addition, Fig. 4 represents the threshold values established for each reference pixel in each image for each class. As it can be seen, in average, the threshold values for the normal, tumor and hypervascularized classes are lower than 10%. However, in the background class there is a higher variability due to the different materials and tissues that can be included in this class.

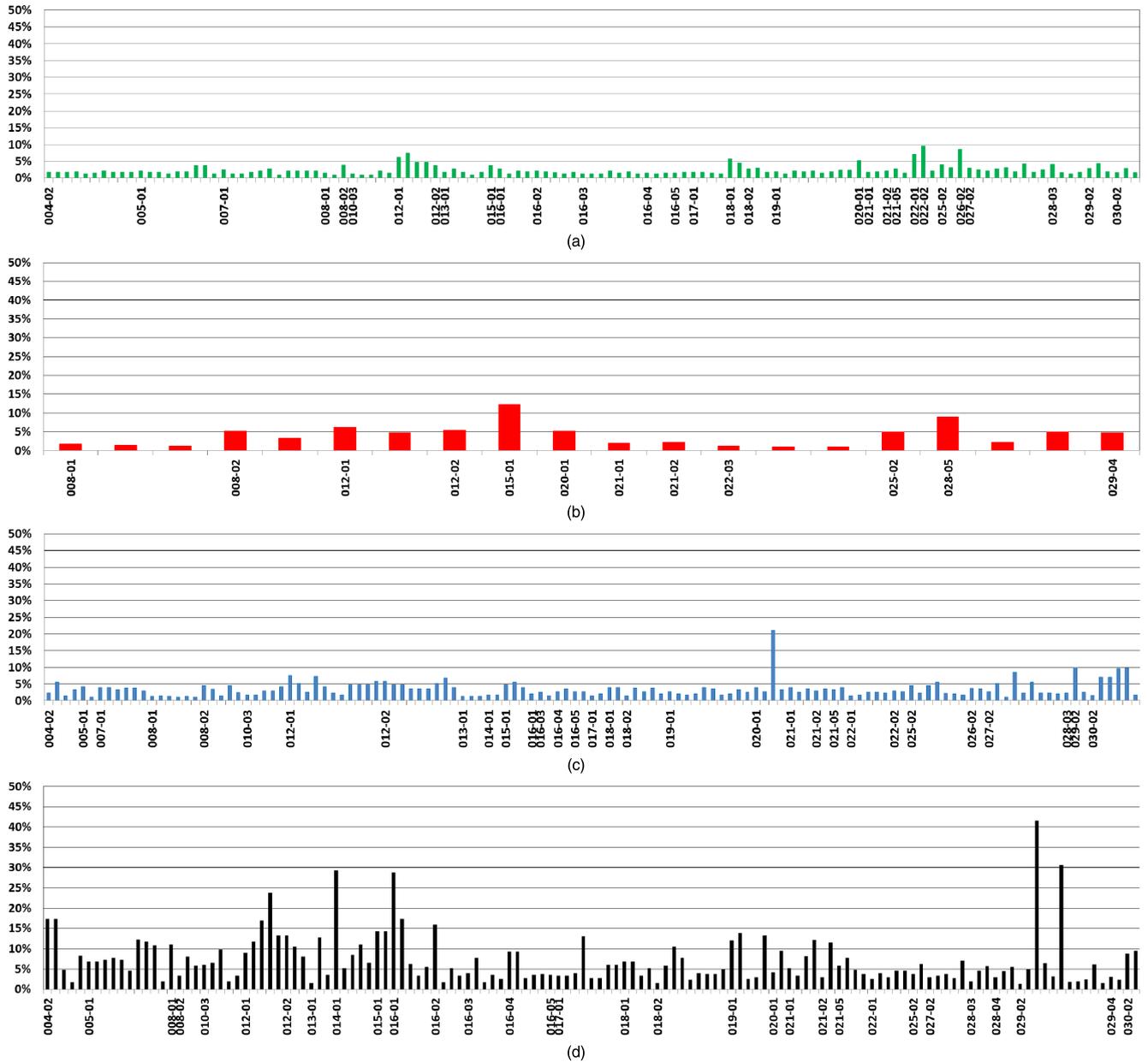

**FIGURE 4.** Threshold values selected for each reference pixel in each image for each class. (a) Normal class. (b) Tumor class. (c) Hypervascularized class. (d) Background class.

## V. REPEATABILITY ANALYSIS OF THE HS ACQUISITION SYSTEM

The developed HS acquisition system was tested by performing two different types of repeatability experiments. The goal of this study was to evaluate the possible sources of systematic errors in the acquisition system and to verify the repeatability of the spectra when images of the same scene were obtained. This section summarizes the procedure and the results acquired in the spatial and spectral repeatability tests performed using three different pairs of HS cubes obtained with the previously introduced system.

### A. REPEATABILITY DATASET

Three different scenes were captured twice and consecutively in the same environmental conditions to perform the repeatability experiments. In total, three pairs of HS cubes were obtained: a white reference tile employed to calibrate the system, a chessboard pattern and a book cover fragment with high spatial and spectral entropy (image with high amount of information). Fig. 5 shows the reconstructed RGB representations of these HS cubes. The white reference tile (Fig. 5.a) and the chessboard pattern (Fig. 5.c) are simple and geometrically uniform. These images were suitable to identify any relevant difference or geometric distortion between each pair of images. The book cover fragment (Fig. 5.e) was selected due to the multiple internal reflections and changes in the reflected light path because of intermediate refractions.

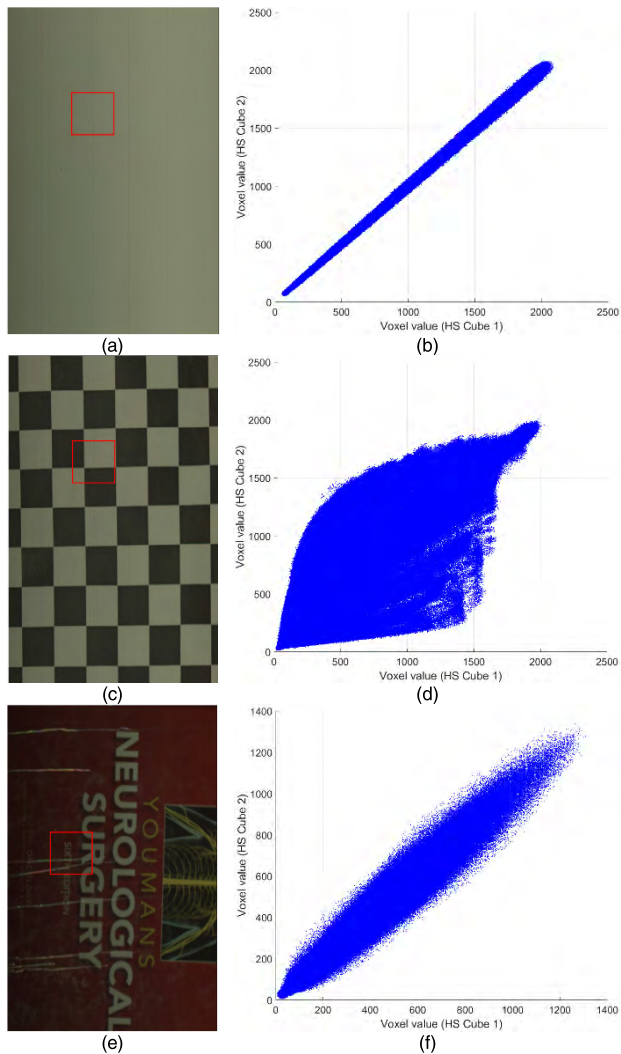

**FIGURE 5.** HS images of the repeatability dataset and their correspondent scatterplot of the voxel values of an example section of 200 × 200 pixels obtained from two HS cubes of the same scene: (a) and (b) White reference tile; (c) and (d) Chessboard pattern; (e) and (f) Book cover fragment.

### B. REPEATABILITY METRICS

The motivation for performing these experiments was to quantify the repeatability of the acquisition system, which enables the analyst to assess the a priori feasibility of performing "correct" regression or classification model for a given target. If the repeatability error is high across all the spectral bands, any statistical modeling will be prone to spurious results [51].

In Fig. 5, the scatterplots of an example section of 200×200 pixels obtained from the two HS cubes of the same target are shown. The selected areas have been surrounded using a red box. In the scatterplot, the voxel values of each HS cube section pair are represented (more than 33 million of voxel pairs). Ideally, this scatterplot should be a straight line, indicating that all the correspondent voxel value pairs contain the same exact information. In this case, the repeatability of the system will be optimal. However, in the real world, the scatterplot will be expanded depending on both the repeatability qualities of the system and the complexity and contrast of the captured scene. In this sense, Fig. 5.b show the scatterplot of the white reference tile (a homogeneous scene), where it is possible to observe that the system offers a low variability in the voxel values from both HS cubes. Nevertheless, when the scene contrast increases (Fig. 5.c), the scatterplot is highly scattered (Fig. 5.d). A small variation in the scene position can produce a high variation in the voxel values due to the high contrast of the chessboard pattern borders. In the scatterplot of the book cover fragment (Fig. 5.f), the expansion is lower than the chessboard pattern due to the scene contrast is lower. This last one is a more realistic example.

The measures and protocols proposed in [51] were employed for these experiments. In [51], the *spatial repeatability* is defined as the differences between the pixel values of the respective spectral layers in two HS cubes from the same scene, while the *spectral repeatability* is defined as the differences between the spectral curves of their respective voxels (value of a pixel in a certain wavelength). The repeatability error (RE) is computed as the root mean square (*rms*) of the differences between two vectors ($x$, $y$) is calculated. The ideal value of RE is zero, but in practice RE is always higher than zero. In this sense, this metric can be modified to represent the noise-to-signal ratio (N/S) or its inverse, signal-to-noise ratio (S/N), being zero and infinite their ideal values. These metrics are defined in (1) and (2), respectively. Furthermore, another metric employed to measure the repeatability of the system is the absolute relative difference percentage (RD) that is defined in (3). In this metric, the relation between the absolute difference and the mean values of the two vectors is computed. In this case, the lower RD, the better that the repeatability of the system will be. S/N and RD metrics were employed to measure the spectral and spatial repeatability of the system by using plot charts, while N/S and RD were employed to represent graphically the spectral repeatability in a certain band.

$$N/S = \frac{rms(x - y)}{rms(x \cup y)} \tag{1}$$

$$S/N = \frac{rms(x \cup y)}{rms(x - y)} \tag{2}$$

$$RD(\%) = \frac{abs(x - y) * 100}{[mean(x) + mean(y)]/2} \tag{3}$$

### C. SPECTRAL REPEATABILITY

Spectral repeatability considers a spatial-slice as a vector by fixing the wavelengths ($\lambda$) and calculating the previously defined metrics. Several HS cubes were created using a pushbroom technique, whereby the complete HS image of the target was generated by joining multiple contiguous non-overlapped image strips (line scans). During each line scan, the CCD sensor acquired a multi-layer slice of the HS cube, where each layer belonged to a predetermined spectral band. Each column of the CCD sensor acquired the spectrum of one pixel and each row (CCD bins rows) contained one spectral band of the HS cube slice. In this sense, it is possible to distinguish between spectral repeatability along the line scans and along the CCD bins rows of the HS cube. Fig. 6 shows an example of how the HS cube is generated using a pushbroom technique.

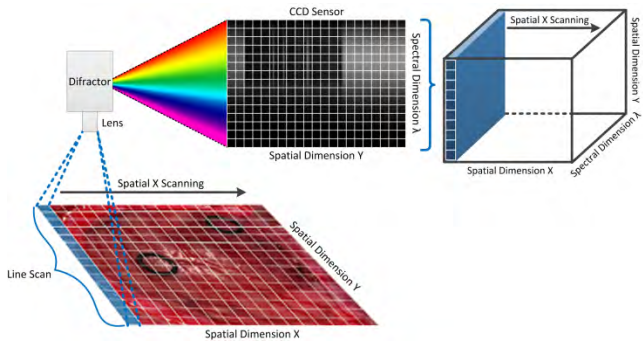

**FIGURE 6.** Pushbroom technique block diagram example.

The spectral repeatability experiments allow measuring possible errors produced due to the vibrations of the acquisition system, the interline scanning errors or errors produced due to differential responses of the CCD bins. These are the most common sources of errors in pushbroom HS systems.

Fig. 7 shows the spectral repeatability results of S/N and RD along the line scans for the three different pairs of HS images. These curves assess the stability of the repeatability indexes from line scan to line scan, as well as detecting the location of possible outlier voxels. As it can be seen in the results, the average of the S/N index decreases when the complexity of the captured scene increases. In the case of the $RD_{mean}$ metric, the mean value of each line scan was calculated and it increases when the complexity of the image increases. Fig. 7.a shows the results obtained for the white reference image, where is possible to observe that when the system captures a spatially homogeneous image, the spectral repeatability along the line scans is stable and minimal. However, in the case of the results obtained for the chessboard pattern image, the peaks obtained in the plot reveal that the repeatability errors are mainly produced in the borders of the image pattern (Fig. 7.b).

In the case of the book cover image (Fig. 7.c) the error is spread in all the line scans and it is mainly produced due to the reflections of the light on the surface and the movement of the system. The worst S/N value is obtained for the chessboard pattern image, which can be caused by the abrupt changes in the pattern due to the motion of the pushbroom sensor across the different scanlines. However, the average $RD_{mean}$ metric is lower than the value obtained for the book cover image.

Fig. 8 presents the spectral repeatability results along the CCD bins for the three pairs of HS images. The dome-shaped curves obtained in the S/N and $RD_{mean}$ metrics of the white reference image (Fig. 8.a) show that the spectral repeatability was better in the center of the CCD array than in the borders. This is mainly caused due to optical aberrations produced at the ends of the sensor and to inhomogeneous illumination across the X pushbroom scan direction. Fig. 8.b and c show the results obtained for the chessboard pattern and the book cover, respectively. In this case, the worst S/N value is obtained for the chessboard pattern image. However, the average $RD_{mean}$ metric is lower than the one obtained for the book cover image.

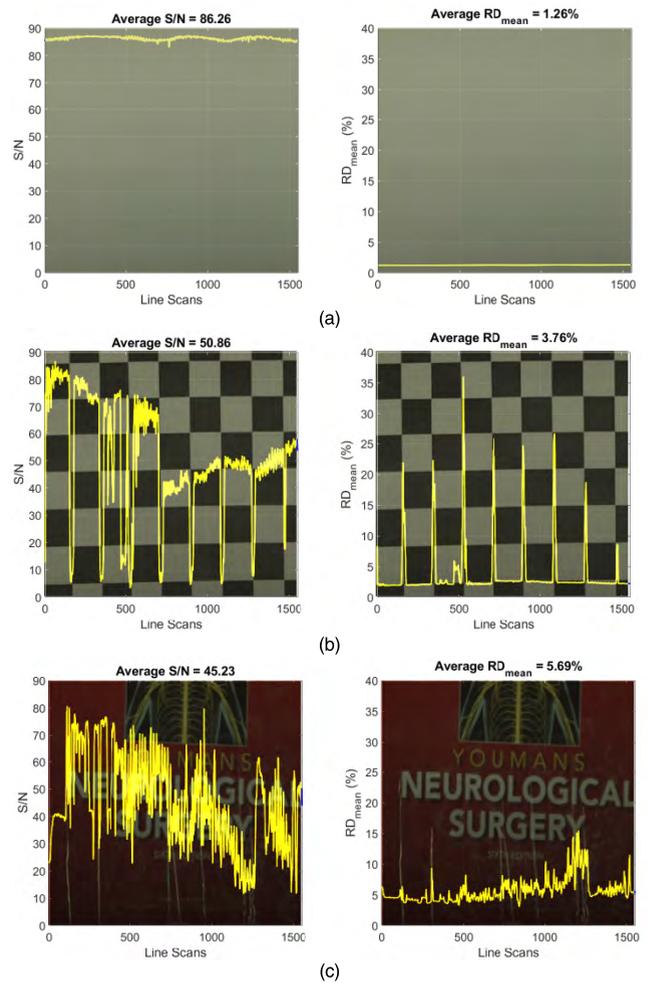

**FIGURE 7.** Spectral repeatability indexes in line scans of the three different HS cube pairs: (a) White reference tile; (b) Chessboard pattern; (c) Book cover fragment.

Finally, Fig. 9 shows the spatial representation of the spectral repeatability indexes of N/S and $RD_{mean}$ of the white reference tile (Fig. 9.a), the chessboard pattern (Fig. 9.b) and the book cover fragment (Fig. 9.c) HS cube pairs at a certain wavelength ($\lambda = 690.78$ nm). This $\lambda$ value was selected as an example to represent the error in a centered wavelength of the spectrum. The repeatability indexes are depicted in false color images. The scales of the false colors on the right side of each image determine the values and ranges of the corresponding repeatability index.

In these false color images, it is possible to identify directly by visual inspection the regions of the images where the repeatability of the system is extremely high or low. In this sense, the repeatability is quite low in the areas of the images with higher spatial entropy. These high-entropy areas cannot be accurately reproduced due to the fine resolution of the spatial scanning required to obtain the HS cube. For example, in the chessboard pattern image (Fig. 9.b) the boundaries of the pattern have a low repeatability while in the book cover image (Fig. 9.c) low repeatability areas are also in the borders of the letters and the drawing as well as in the relief produced by plasticizing the book cover. The spectral repeatability is a function of the ambient light and the nature of the materials presented in the scene [51].

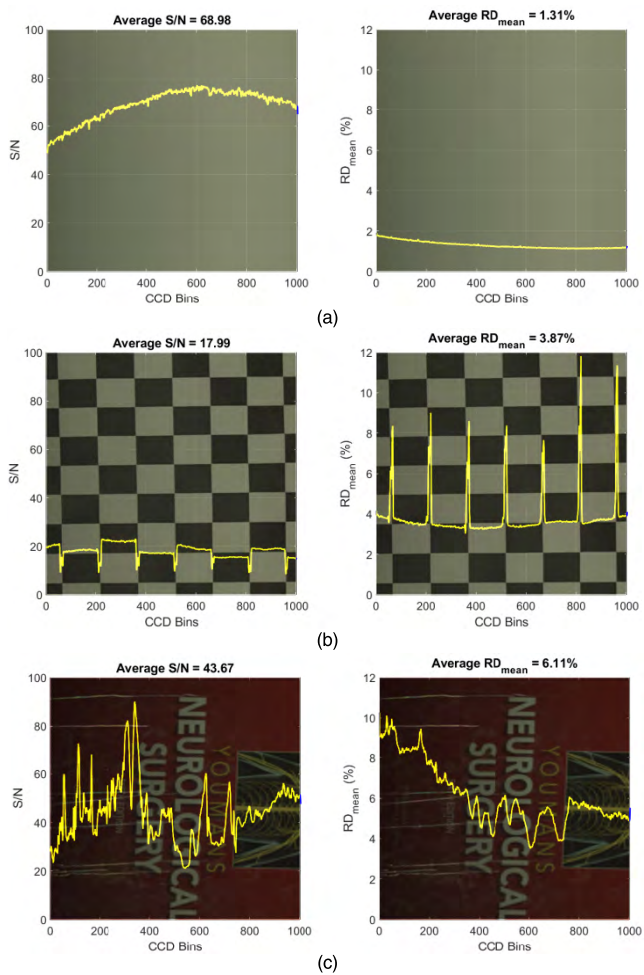

**FIGURE 8.** Spectral repeatability indexes in CCD bins of the three different HS cube pairs: (a) White reference tile; (b) Chessboard pattern; (c) Book cover fragment.

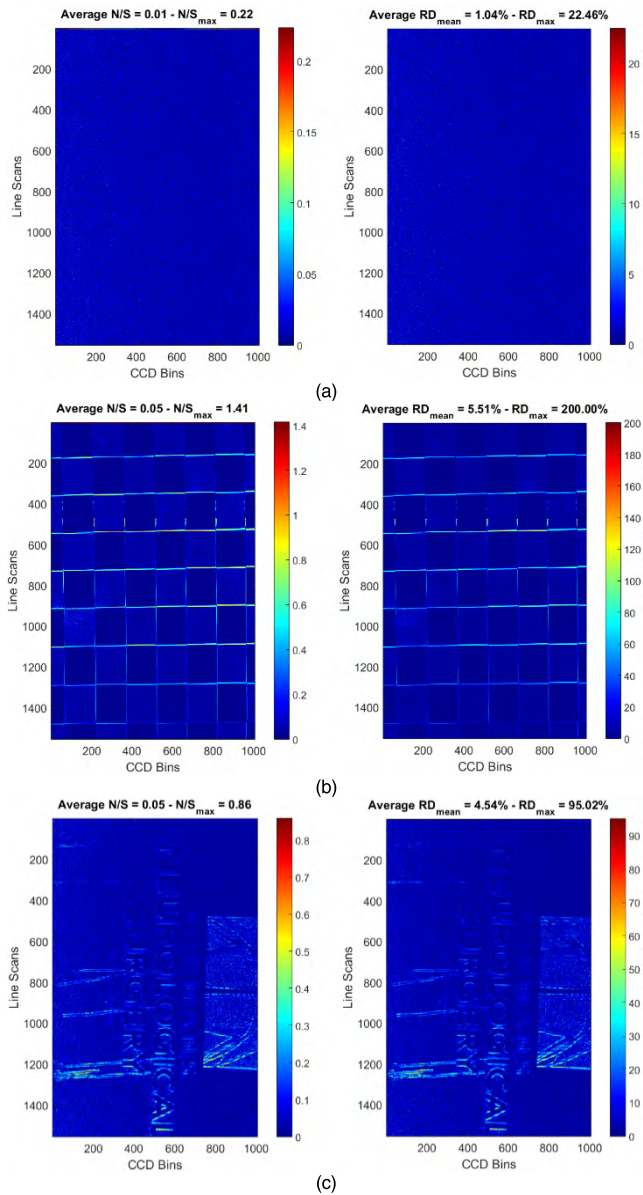

**FIGURE 9.** Spatial representation of the spectral repeatability indexes of N/S and RD$_{mean}$ of the three different HS cube pairs at a certain wavelength ($\lambda$ = 690.78 nm): (a) White reference tile; (b) Chessboard pattern; (c) Book cover fragment.

### D. SPATIAL REPEATABILITY

The spatial repeatability measures the differences between pixel values at a certain wavelength in HS cubes of the same scene. Fig. 10 shows the S/N and RD$_{mean}$ metrics for each one of the HS cube pairs to measure the spatial repeatability of the HS acquisition system. As it can be seen in the results, the spatial repeatability of the system was better in the center wavelengths and worse towards the first and the last spectral bands. This is produced because in the shorter wavelengths the photons have higher energy than in the longer ones. Consequently, they are absorbed closely to the CCD sensor surface and not by the active part of the detector [51]. The oscillations presented in the S/N plot of the book cover fragment (Fig. 10.c) are produced due to the presence of multiple materials that have low and high reflectance values in different wavelengths.

Taking into account the results of these experiments, the optimal spectral range where the system can operate efficiently is comprised between 450 and 900 nm approximately. For this reason, during the pre-processing of the HS cubes, the reflectance values outside of this range were avoided.

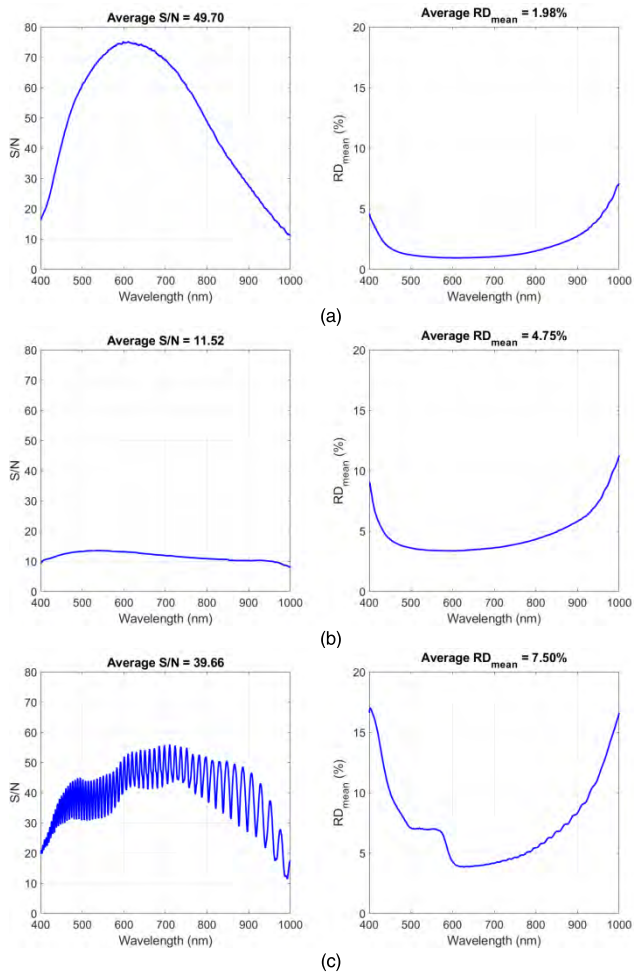

**FIGURE 10.** Spatial repeatability indexes of the three different HS cube pairs: (a) White reference tile; (b) Chessboard pattern; (c) Book cover fragment.

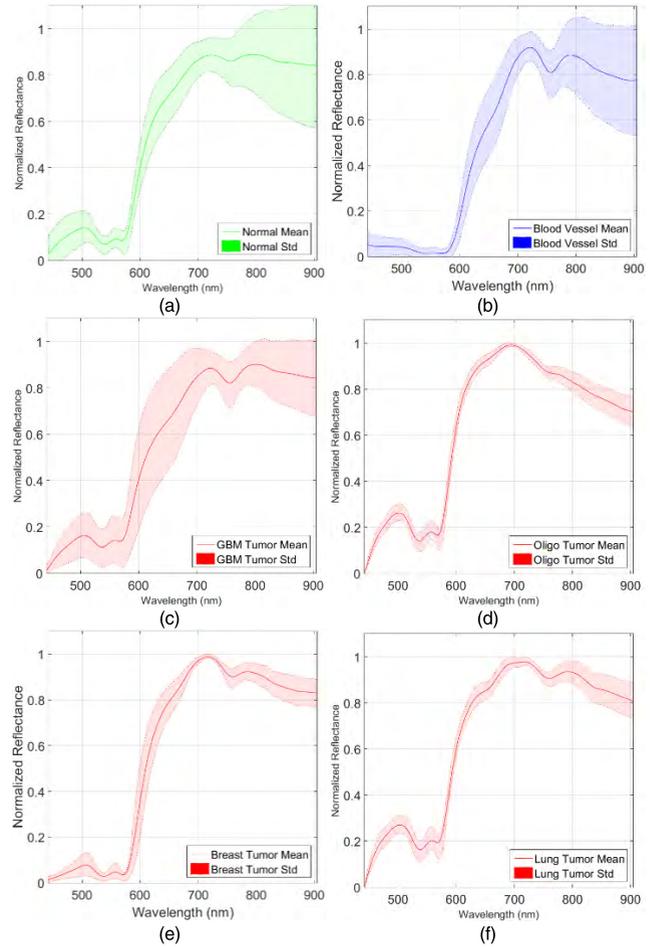

**FIGURE 11.** Average and standard deviation (Std) across all the patients of the different types of spectral signatures in the labeled database: (a) Normal tissue; (b) Blood vessels; (c) GBM tumor; (d) Anaplastic oligodendroglioma tumor; (e) Secondary breast tumor; (f) Secondary lung tumor.

## VI. DATABASE SUMMARY

The gold standard dataset generated using the previously presented labeling methodology is composed by 36 *in-vivo* brain surface images from 22 different patients. In this images, tumor and normal tissue were labeled (when possible) as well as the blood vessels and other tissues, materials or substances that can be presented in the surgical scene and are not relevant for the surgical resection procedure (called hereafter background). Fig. 11 shows the average and standard deviation of each type of tissue presented in the labeled dataset across all the patients, differentiating between the tumor types. This dataset includes both primary (grade IV glioblastoma and grade III and II anaplastic oligondendrogliomas) and secondary (lung and breast) tumors. Fig. 12 shows the RGB representations of the HS cubes that compose the *in-vivo* HS human brain image database, where the approximate tumor area is surrounded by a yellow line in order to identify the rubber ring marker that corresponds to the pathological analysis performed over the tumor tissue. The codes above each image are the Patient ID and Image ID shown in Table 1, which summarizes the total number of images available in this database, its characteristics and the pathological diagnosis of each image. Furthermore, the total number of labeled pixels for each class and image are detailed. As it can be seen, some images were diagnosed as a certain type of tumor; however, no tumor samples were labeled due to the difficulties to perform this procedure (image perspective, deep location of the tumor, etc.).

## VII. LIMITATIONS

Due to the difficulties to acquire in-vivo HS images during human neurosurgical procedures, during 2 years we were only able to capture the most common tumor types that were available during the surgical procedures in both participant hospitals (mainly GBM tumors). Since the customized HS acquisition system is a preliminary demonstrator, the system has several limitations. One of them is that the system was only able to capture tumor images where the tumor was either on the surface or in a deeper layer, but easy enough to be focused and captured. Moreover, the HS pushbroom camera employed for the demonstrator development requires to perform a spatial scanning, considerably increasing the

**TABLE 1.** Summary of the *in-vivo* HS human brain image database.

| Hospital | Patient ID | Image ID | Size (height x width x bands) | #Labeled Pixels (#Reference Pixels) | | | | Diagnosis |
|---|---|---|---|---|---|---|---|---|
| | | | | N | T | BV | B | |
| UHDRN | 004 | 02 | 389 x 345 x 826 | 5,007 (1) | 0 | 965 (4) | 1,992 (5) | Normal Brain |
| | 005 | 01 | 483 x 488 x 826 | 6,061 (9) | 0 | 1,727 (2) | 20,483 (10) | Renal Carcinoma (S) |
| | 007 | 01 | 582 x 400 x 826 | 7,714 (11) | 0 | 1,089 (6) | 0 | Normal Brain |
| | 008 | 01 | 460 x 549 x 826 | 2,295 (2) | 1,221 (3) | 1,331 (6) | 630 (1) | Grade IV Glioblastoma (P) |
| | | 02 | 480 x 553 x 826 | 2,187 (1) | 138 (2) | 1,000 (5) | 7,444 (2) | Grade IV Glioblastoma (P) |
| | 010 | 03 | 371 x 461 x 826 | 10,626 (5) | 0 | 2,332 (5) | 3,972 (5) | Grade IV Glioblastoma (P) |
| | 012 | 01 | 443 x 497 x 826 | 4,516 (4) | 855 (2) | 8,697 (11) | 1,685 (5) | Grade IV Glioblastoma (P) |
| | | 02 | 445 x 498 x 826 | 6,553 (1) | 3,139 (1) | 6,041 (9) | 8,731 (3) | Grade IV Glioblastoma (P) |
| | 013 | 01 | 298 x 253 x 826 | 1,827 (5) | 0 | 129 (3) | 589 (3) | Lung Carcinoma (S) |
| | 014 | 01 | 317 x 244 x 826 | 0 | 30 (1) | 64 (2) | 1,866 (5) | Grade IV Glioblastoma (P) |
| | 015 | 01 | 376 x 494 x 826 | 1,251 (1) | 2,046 (1) | 4,089 (3) | 696 (2) | Grade IV Glioblastoma (P) |
| | 016 | 01 | 335 x 323 x 826 | 3,970 (4) | 0 | 246 (1) | 12,002 (5) | Normal Brain |
| | | 02 | 335 x 326 x 826 | 349 (5) | 0 | 0 | 2,767 (4) | Normal Brain |
| | | 03 | 315 x 321 x 826 | 603 (7) | 0 | 234 (2) | 1,696 (5) | Normal Brain |
| | | 04 | 383 x 297 x 826 | 1,178 (3) | 0 | 1,064 (2) | 956 (5) | Grade IV Glioblastoma (P) |
| | | 05 | 414 x 292 x 826 | 2,643 (2) | 0 | 452 (2) | 5,125 (1) | Grade IV Glioblastoma (P) |
| | 017 | 01 | 441 x 399 x 826 | 1,328 (4) | 0 | 68 (2) | 3,069 (8) | Grade IV Glioblastoma (P) |
| | 018 | 01 | 479 x 462 x 826 | 13,450 (2) | 0 | 488 (2) | 9,773 (4) | Grade I Ganglioglioma (P) |
| | | 02 | 510 x 434 x 826 | 4,813 (2) | 0 | 958 (5) | 5,895 (9) | Grade I Ganglioglioma (P) |
| | 019 | 01 | 601 x 535 x 826 | 6,499 (9) | 0 | 1,350 (10) | 1,933 (5) | Meningioma |
| | 020 | 01 | 378 x 330 x 826 | 1,842 (1) | 3,655 (1) | 1,513 (20) | 2,625 (2) | Grade IV Glioblastoma (P) |
| | 021 | 01 | 452 x 334 x 826 | 3,405 (2) | 167 (1) | 793 (3) | 5,330 (4) | Breast Carcinoma (S) |
| | | 02 | 448 x 324 x 826 | 2,353 (1) | 31 (1) | 555 (2) | 2,137 (2) | Breast Carcinoma (S) |
| | | 05 | 433 x 340 x 826 | 969 (2) | 0 | 1,637 (2) | 1,393 (4) | Breast Carcinoma (S) |
| | 022 | 01 | 597 x 527 x 826 | 2,806 (1) | 0 | 1,064 (5) | 3,677 (5) | Grade III Anaplastic Oligodendroglioma (P) |
| | | 02 | 611 x 527 x 826 | 8,174 (2) | 0 | 680 (2) | 0 | Grade III Anaplastic Oligodendroglioma (P) |
| | | 03 | 592 x 471 x 826 | 0 | 96 (3) | 0 | 0 | Grade III Anaplastic Oligodendroglioma (P) |
| UHS | 025 | 02 | 473 x 403 x 826 | 977 (2) | 1,282 (1) | 907 (7) | 3,687 (2) | Grade IV Glioblastoma (P) |
| | 026 | 02 | 340 x 324 x 826 | 507 (1) | 0 | 128 (2) | 0 | Normal Brain |
| | 027 | 02 | 493 x 476 x 826 | 6,352 (9) | 0 | 5,606 (9) | 21,785 (5) | Normal Brain |
| | 028 | 03 | 422 x 398 x 826 | 2,839 (4) | 0 | 73 (1) | 13,341 (3) | Normal Brain |
| | | 04 | 482 x 408 x 826 | 0 | 0 | 0 | 10,025 (3) | Lung Adenocarcinoma (S) |
| | | 05 | 482 x 390 x 826 | 0 | 1,920 (3) | 0 | 0 | Lung Adenocarcinoma (S) |
| | 029 | 02 | 365 x 371 x 826 | 2,098 (3) | 0 | 3,341 (2) | 11,258 (11) | Normal Brain |
| | | 04 | 399 x 342 x 826 | 0 | 1,748 (1) | 0 | 3,785 (2) | Grade II Anaplastic Oligodendroglioma (P) |
| | 030 | 02 | 382 x 285 x 826 | 2,050 (3) | 0 | 9,242 (6) | 15,337 (2) | Normal Brain |
| Total | 22 Operations - 36 Captures | | | 117,242 (118) | 16,328 (20) | 57,863 (127) | 185,684 (132) | |

*(N) Normal tissue; (T) Tumor tissue; (BV) Blood vessel; (B) Background; (S) Secondary; (P) Primary.

acquisition time of the HS image. The inherit movement of the undergoing patient's brain and also the possible artifacts, which can appear in the image (such as extravasated blood or surgical serum) during the acquisition, can affect the spatial coherence of the image. In this sense, snapshot cameras (HS cameras that are able to acquire both the spatial and spectral features of a scene in a single shot) are the most suitable option for this application, achieving real-time image acquisition. However, the number of spectral bands in these cameras is extremely lower compared to pushbroom cameras (~10 times less). Therefore, further investigations must be performed by using HS images captured with pushbroom cameras (with high spectral resolution) in order to identify the most relevant bands that allow the identification and delineation of brain tumors. These results will provide valuable feedback to snapshot camera manufacturers that will be able to develop specific HS sensors for this particular application.

Due to the limited availability of patients, it is an extremely hard task to achieve a comprehensive database that covers the entire heterogeneity that could be found across the different patients and the different types of brain tumors. For this reason, in this paper we present a preliminary database that can be employed to perform further investigations in the use of HSI to: 1) identify the different tissue and tumor types; 2) delineate the boundaries of the tumors, parenchyma, blood vessels or other regions of interest for the neurosurgeons; 3) provide other kind of information to the surgeons that can be useful to improve the outcomes of the surgery.

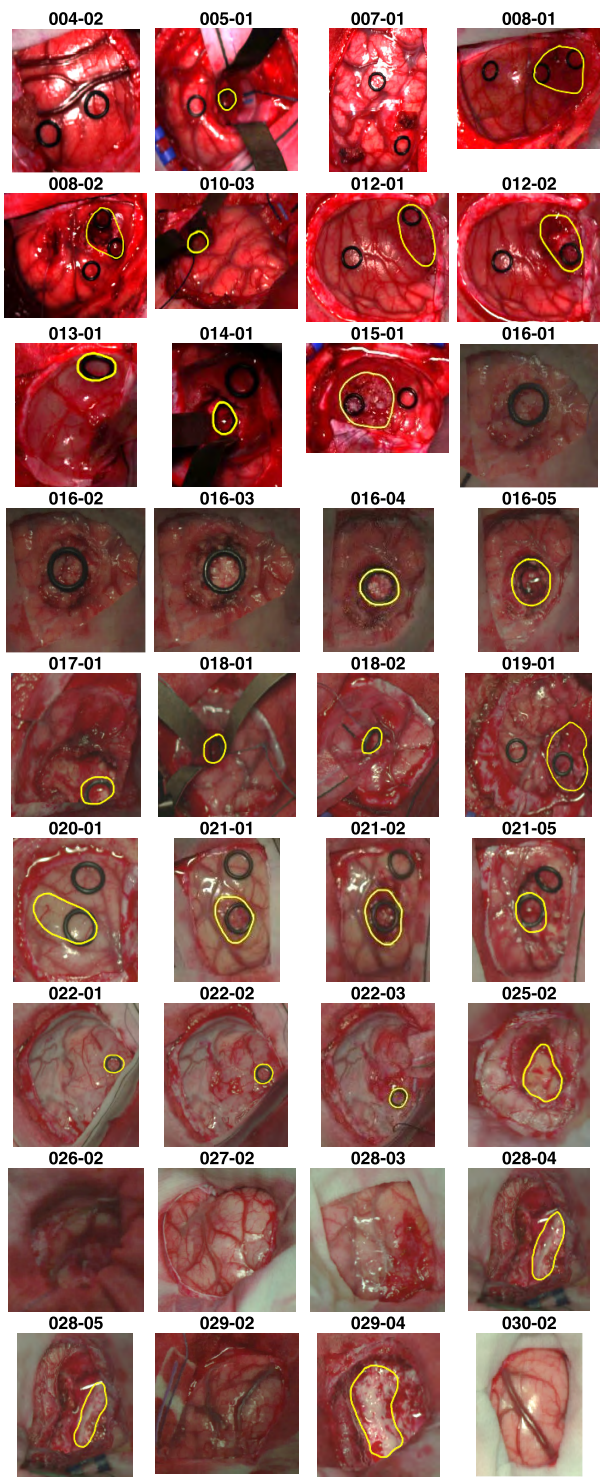

**FIGURE 12.** RGB representations of the HS images that compose the in-vivo HS human brain image database. The numeric code shown above each image represents the Patient ID and Image ID (PatientID-ImageID) that are detailed in Table 1. Yellow lines surround the approximate tumor area identified by the neurosurgeon experience and the indications of the IGS system.

The methodology presented in this paper to generate a labeled dataset that can be used for training supervised classifiers, mainly exploits the spectral characteristics of HSI to differentiate between the different types of tissues. Intraoperative MRI and the knowledge and experience of the surgeons that participated in the project were employed as a guidance tool to identify the suitable areas to be labeled in the images. In addition, the pathological analysis results of the tumor tissue resected during the surgical procedures were employed to accurately identify the tumor areas and type. In future data acquisition campaigns, it is expected that more types of tumors will be included in the database with a more detailed pathological description.

Moreover, further investigations should be done in order to properly correlate the biological properties of the different brain tissue types (especially of the different types of tumors and normal tissues) with the spectral responses obtained by the HS cameras at each wavelength. This study could help in the better discrimination of the different types of tissues, providing better outcomes in the delineation and identification of the tumors, especially in the surrounding normal tissue infiltrated by the tumor. Specific studies should be done to determine these differences and provide a more detailed labeled dataset. Although the labeled dataset presented in this work was obtained conservatively (we tried to label only the pixels where the experts were highly confident that belonged to a certain class) using the methodology based on the combination between the SAM algorithm and the pathological analysis, a new detailed labeled dataset could be used to validate it. In addition, other types of neurosurgical procedures not involving brain tumor should be included in further data campaigns in order to obtain a higher number of normal tissue samples, ensuring that they are not affected by the infiltrated tumor.

## VIII. CONCLUSIONS

The work performed during the execution of the HELICoiD European project, developing an intraoperative system capable of acquiring HS images during neurosurgery procedures, allows giving to the international scientific community open access to the first public in-vivo hyperspectral human brain image database specifically for brain cancer detection. We believe that the public access of this database could promote advances in the field of in-vivo brain research based on hyperspectral imaging, not only for brain cancer research but also for other brain problems such as tissue oxygenation identification, hemoglobin peaks identification in the spectral signatures, parenchymal area identification, brain vascularization map generation, etc.

The HS acquisition system was assessed during surgical procedures in two different hospitals in the UK and Spain, obtaining 36 images from 22 different patients in the VNIR spectral range (between 400 and 1000 nm). The obtained HS images were labeled by specialists (neurosurgeons) in order to offer a highly reliable four-classes labeled dataset, where it was possible to distinguish between normal brain tissue, brain tissue affected by cancer (both primary and secondary), blood vessels and background elements. In this paper, the repeatability experiments (both spectral and spatial) performed to the HS acquisition system in order to assess the quality of the obtained data have been presented. In addition, this repeatability analysis revealed that the suitable spectral range where the system can operate more efficiently is comprised between 450 and 900 nm. This database has been successfully employed to generate classification maps where the tumor boundaries are identified and delineated by using traditional machine learning techniques [32], [52] and also deep learning approaches [53], [54]. In addition, it has been employed as input images for the acceleration of these HSI processing algorithms obtaining processing surgical-time results that range between 15 to 70 seconds, depending on the employed platform for the acceleration [55]–[60]. All the HS image files generated from this study are available from: http://hsibraindatabase.iuma.ulpgc.es.

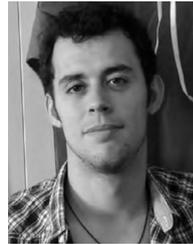

**HIMAR FABELO** received the degree in telecommunication engineering and the master's degree in telecommunication technologies from the University of Las Palmas de Gran Canaria, Las Palmas de Gran Canaria, Spain, in 2013, and 2014, respectively. Since 2014, he has been conducting his research activity in the Integrated System Design Division, Institute for Applied Microelectronics, University of Las Palmas de Gran Canaria, in electronic and bioengineering. In 2015, he started to work as a Coordination Assistant and a Researcher with the HELICoiD European Project, co-funded by the European Commission. In 2018, he performs a research stay in the Department of Bioengineering, Erik Jonsson School of Engineering and Computer Science, The University of Texas at Dallas, collaborating with Prof. B. Fei in the use of medical hyperspectral imaging analysis using deep learning. His research interests include the use of machine learning and deep learning techniques applied to hyperspectral images to discriminate between healthy and tumor samples for human brain tissues in real time during neurosurgical operations.

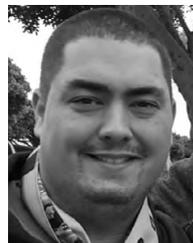

**SAMUEL ORTEGA** received the degree in telecommunication engineering and the master's (research) degree in telecommunication technologies from the University of Las Palmas de Gran Canaria, Spain, in 2014, and 2015, respectively. Since 2015, he has been conducting his research activity in the Integrated System Design Division, Institute for Applied Microelectronics, University of Las Palmas de Gran Canaria, in electronic and bioengineering. In 2015, he started to work as a Coordination Assistant and a Researcher with the HELICoiD European Project, co-funded by the European Commission. His current research interest includes the use of machine learning algorithms in medical applications using hyperspectral images.

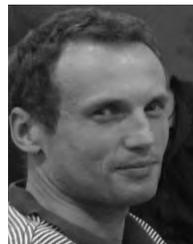

**ADAM SZOLNA** received the M.D. and Ph.D. degrees from the Medical University of Lodz, Poland, in 1998 and 2005, respectively, as a result of a five year clinical study conducted on the patients operated for dystonia. He is currently a Specialist in neurosurgery. He is also a member of the permanent staff of the Neurosurgery Department, University Hospital Doctor Negrin, Las Palmas de Gran Canaria. He has co-authored a book Functional Neurosurgery. His current research interests include hyperspectral imaging and real-time intraoperative brain tumour detection, participating in the HELICOID Project as a Principal Investigator. He is a member of the Spanish Society of Neurosurgeons, the European Society for Stereotactic and Functional Neurosurgery, the Section of Stereotaxy and Functional Neurosurgery, Polish Society of Neurosurgeons, and the Polish Society of Neurosurgeons. He was serving 16 years for Military Health Service, Poland, and NATO, up to the rank of major.

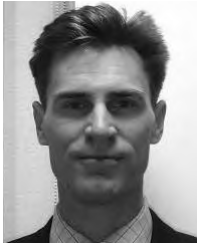

**DIEDERIK BULTERS** is currently a Consultant Neurosurgeon and an Honorary Senior Clinical Lecturer with the University of Southampton and trained in Edinburgh, Southampton, and Cambridge. The team received grants from EPSRC, TSB, NIHR, EU, and Wessex Medical Research. His current research interests include neurovascular disease, neurotrauma, and neuro-oncology.

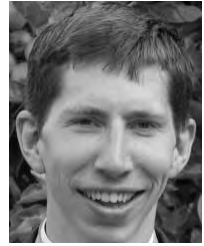

**HARRY BULSTRODE** is currently a Neurosurgery Registrar and a Wellcome Research Training Fellow. He has trained in Cambridge, Oxford, London, Southampton, and Edinburgh. His research interests include molecular biology of neural stem cells and brain tumours, and to the application of spectral analysis to guide brain tumour resection.

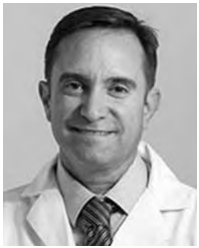

**JUAN F. PIÑEIRO** received the M.D. degree from the High Institute of Medical Sciences of Villa Clara, Cuba, in 1993. He is currently a Specialist in neurosurgery. He is also the former Head of the Neurosurgical Department, Dr. Gustavo Aldereguia Hospital, Cienfuegos, Cuba. He is also a Specialist in neurosurgery with University Hospital Doctor Negrin, Las Palmas de Gran Canaria, and also with Vithas Santa Catalina Hospital, Las Palmas de Gran Canaria. He has authored a book chapter and several papers and contributions to technical conferences. His research interests include head trauma, neurovascular surgery, and the application of new technologies for oncological Neurosurgery. He is a member of the Spanish Society of Neurosurgery, the Cuban Society of Neurology and Neurosurgery, and the European Association of Neurological Surgeons.

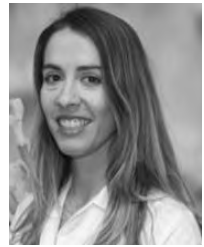

**SARA BISSHOPP** received the M.D. degree from the Medicine High School, La Laguna University, Spain, in 2009, and the Honors degree. She has performed a 3- month's rotation internship in the Pediatrics Unit of Hospital Virgen del Rocio, Seville. She received the training in spine surgery with a 3-months rotation internship in the Spine Unit, Virgen del Rocío Hospital. Because of her interest in oncological Neurosurgical research, she has participated in the HELICOID Project as a Co-Investigator. She is currently a Specialist in neurosurgery. She is also a member of the Staff of the Neurosurgery Department, University Hospital Doctor Negrin, Las Palmas de Gran Canaria. She is a member of the Spanish Society of Neurosurgery and the European Society of Neurosurgery.

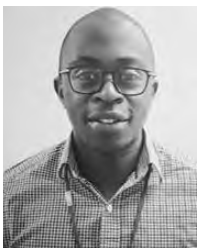

**SILVESTER KABWAMA** is currently a Neurosurgery Research Fellow with the Wessex Neurological Centre. He trained at the University of Southampton and King's College London. He is responsible for the day to day running of the HELI-CoiD Project at the Wessex Neurological Centre. His research interests include the development and management of brain tumours.

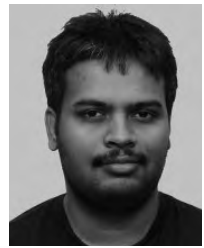

**B. RAVI KIRAN** received the Ph.D. degree from University Paris-Est, in 2014 on optimization over energetic lattices for image segmentation. He was a Postdoctoral Researcher with the CAOR Lab, Mines ParisTech, in 2015, on hyperspectral image processing for tumor delineation, on which he continues to collaborate with the Helicoid team to this day. He was also a Postdoctoral Researcher with the DATA lab, ENS Paris, in 2016, on unsupervised streaming time series anomaly detection. He was a temporary Assistant Professor and a Research with University Lille 3, from 2016 to 2017, during which he primarily taught machine learning and data mining. Since 2018, he has been a Research Engineer with Autonomous Systems, AKKA Technologies, where he works on deep reinforcement learning for autonomous driving.

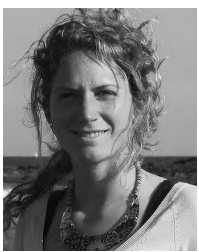

**ARUMA J-O'SHANAHAN** is currently pursuing the Ph.D. degree with the University of La Laguna, Tenerife, Spain. She received the training with a fellowship in vascular neurosurgery with the University Hospital of Helsinki, Finland, along with observerships at Teishinkai Hospital, Japan, and also with Toronto Western Hospital, Canada. She is currently a Specialist in neurosurgery. She is also a member of the Staff of the Neurosurgery Department, the University Hospital Doctor Negrin, Las Palmas de Gran Canaria, where she is actually a part of the Neuro Interventionist Radiology Team. Her current research interests include vascular neurosurgery. She is a member of the College of Physicians of Las Palmas de Gran Canaria, the Spanish Society of Neurosurgeons, and the European Society of Neurosurgery.

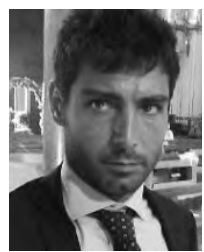

**DANIELE RAVÌ** received the master's degree (summa cum laude) in computer science from the University of Catania, in 2007, and the Ph.D. degree with the Department of Mathematics and Computer Science, University of Catania, Italy, in 2014. From 2008 to 2010, he was with STMicroelectronics (Advanced System Technology Imaging Group) as a Consultant. He was with the Centre for Vision, Speech and Signal Processing, University of Surrey, U.K. Since 2014, he has been a Research Associate with the Hamlyn Centre for Robotic Surgery, Imperial College London, for almost four years. He is currently a Senior Research Associate in computer vision, machine learning, image-guided surgery and smart sensing with the Wellcome/EPSRC Centre Interventional and Surgical Sciences, University College of London. He has contributed to several research projects funded by the EU and the industry. He has co-authored different papers including book chapters, international journals, and international conference proceedings.

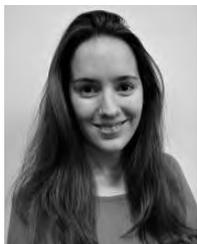

**RAQUEL LAZCANO** received the B.Sc. degree in communication electronics engineering and the M.Sc. degree in systems and services engineering for the information society from the Universidad Politécnica de Madrid, Spain, in 2014 and 2015, respectively, where she is currently pursuing the Ph.D. degree in systems and services engineering for the information society with the Electronic and Microelectronic Design Group. In 2015, she stayed four months at the Institute of Electronics and Telecommunications of Rennes (IETR), National Institute of Applied Sciences (INSA), France, as an Interchange Student of the M.Sc. degree. She has authored or co-authored five indexed journals and 16 contributions to technical conferences. Her research interests include high-performance multicore processing systems, real-time hyperspectral image processing, and the automatic optimization of the parallelism in real-time systems.

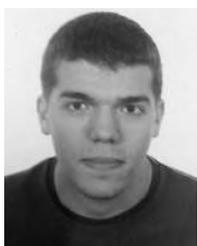

**DANIEL MADROÑAL** received the B.Sc. degree in communication electronics engineering and the M.Sc. degree in systems and services engineering for the information society from the Universidad Politécnica de Madrid, Spain, in 2014 and 2015, respectively, where he is currently pursuing the Ph.D. degree in systems and services engineering for the information society with the Electronic and Microelectronic Design Group (GDEM). In 2015, he stayed 4 months at the National Institute of Applied Sciences (INSA), France, as an Interchange Student of the M.Sc. degree. He has authored or co-authored five indexed journals and 16 contributions to technical conferences. His research interests include high-performance multicore processing systems, real-time hyperspectral image processing, and the automatic optimization of the energy consumption in high-performance systems.

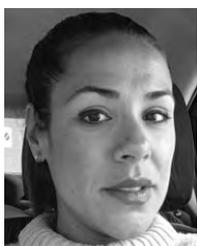

**CORALIA SOSA** received the M.D. and Ph.D. degrees from the Medical Faculty, University of Las Palmas de Gran Canaria, in 2007 and 2016, respectively, as a result of a clinical study on the patients subaracnoid haemorrhage in poor grade neurological status. She is currently a Specialist in neurosurgery. She is also a member of the Staff of the Neurosurgery Department, University Hospital Doctor Negrin, Las Palmas de Gran Canaria. She trained for four months at the Addembrooke Hospital, Cambridge, in brain cancer surgery. Her current research interests include hyperspectral imaging and real-time intraoperative brain tumour detection, participating in the HELICOID project as a Principal Investigator. She is a member of Spanish and European Society of Neurosurgeons.

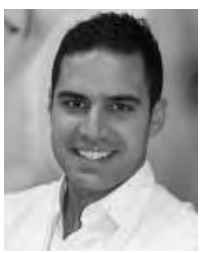

**CARLOS ESPINO** is currently a Specialist in neurosurgery. He is also a member of the permanent staff of the Neurosurgery Department, University Hospital Doctor Negrin, Las Palmas de Gran Canaria.

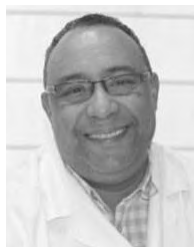

**MARIANO MARQUEZ** is currently a Specialist in neurosurgery. He is also a member of the permanent staff of the Neurosurgery Department, University Hospital Doctor Negrin, Las Palmas de Gran Canaria.

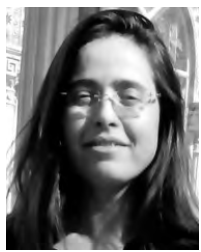

**MARÍA DE LA LUZ PLAZA** received the Ph.D. degree by ULPGC, in 2002, where she is currently Associate Professor of histopathology with the Faculty of Medicine. She is also a Specialist in histopathology with the Department of Histopathology, University Hospital Doctor Negrin, Las Palmas de Gran Canaria. She is a member of the Spanish Society of Pathologic Anatomy.

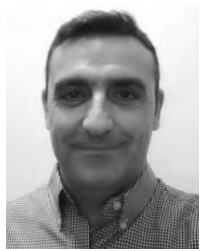

**RAFAEL CAMACHO** is currently a Specialist in pathology. He is also the former Head of the Pathology Department, University Hospital Doctor Negrin, Las Palmas de Gran Canaria, where he is also a General Pathologist and especially as a Nephropathologist. He is also the Head of the Department of Electron Microscopy, Pathology Department. He is a member of the European Nephropatology Working Group, College of Physicians of Las Palmas, and the Spanish Society of Pathologists.

**DAVID CARRERA** graduated from the Universidad Autónoma de Barcelona, Barcelona, in 2006. After completing neurology training at the Neurology Department, Hospital Universitario Germans Trias i Pujol, Badalona, in 2011, he undertook an observership at the Neurocritical Care Unit, Columbia University Medical Center, New York, and, after, a fellowship at the Stroke Unit, Hospital de la Santa Creu i Sant Pau, Barcelona. In 2014, he started his neurosurgery training, and he is in the fourth year of the residency program. He is currently a Neurosurgery Trainee with the Neurosurgery Department, University Hospital Doctor Negrin, Las Palmas de Gran Canaria. He has published peer-reviewed articles in both national and international journals.

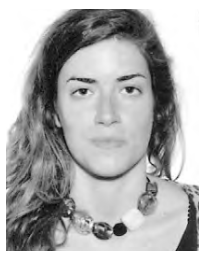

**MARÍA HERNÁNDEZ** is currently a Medical Trainee in neurosurgery. After finishing her fifth and last year of training, she is currently with the Department of Neurosurgery, Hospital University Hospital Doctor Negrin. She has performed a 3-months rotation internship in the Pediatric Unit, Hospital Virgen del Rocío, Seville. She is currently performing a 2-months observership in the Neurooncology Unit, Addenbrookes Hospital, University of Cambridge. She is also a member of the Spanish Society of Neurosurgery.

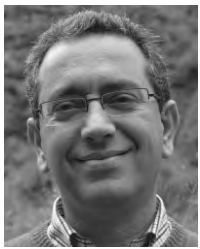

**GUSTAVO M. CALLICÓ** (M'08) received the M.S. degree (Hons.) in telecommunication engineering and the Ph.D. and European Doctorate degrees (Hons.) from the University of Las Palmas de Gran Canaria, in 1995 and 2003, respectively. From 1996 to 1997, he was granted with a research grant from the National Educational Ministry and, in 1997, he was hired by the university as an Electronic Lecturer. In 1994, he joined the Institute for Applied Microelectronics (IUMA) and, from 2000 to 2001, he stayed at Philips Research Laboratories, Eindhoven, The Netherlands, as a Visiting Scientist, where he developed his Ph.D. thesis. He currently develops his research activities in the Integrated Systems Design Division, IUMA. He is currently an Associate Professor with ULPGC.

Since 2015, he has been responsible for the scientific-technological equipment project called Hyperspectral image acquisition system of high spatial and spectral definition, granted by the General Directorate of research and management of the National R&D Plan, funded through the General Directorate of Scientific Infrastructure. He has been the Coordinator of the European Project HELICoiD (FET, Future, and Emerging Technologies) under the Seventh Framework Program. He has more than 110 publications in national and international journals, conferences, and book chapters. He has participated in 18 research projects funded by the European Community, the Spanish Government, and international private industries. His current research interests include hyperspectral imaging for real-time cancer detection, real-time super-resolution algorithms, synthesis-based design for SOCs and circuits for multimedia processing and video coding standards, especially for H.264 and SVC. He has been an Associate Editor of the IEEE TRANSACTIONS ON CONSUMER ELECTRONICS, since 2009, where he is currently a Senior Associate Editor. He has been an Associate Editor of the IEEE ACCESS, since 2016.

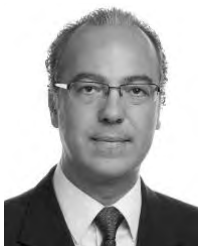

**JESÚS MORERA MOLINA** received the master's degree in healthcare management from the University of Pompeu Fabra. He has been the Deputy Director of the University Hospital Doctor Negrin, and a member of the Board of the College of Physicians of Las Palmas and of the National Commission on Neurosurgery. He is currently a Specialist in neurosurgery. He is currently the Head of neurosurgery with University Hospital Doctor Negrin, Las Palmas de Gran Canaria, and also a Lecturer with the University of Las Palmas de Gran Canaria. He is a member of the Parliament of the Canary Islands.

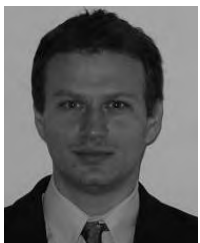

**BOGDAN STANCIULESCU** received the B.Sc. degree in applied physics and the M.Sc. degree in electronics from the University of Bucharest, and the Ph.D. thesis in computer vision. His thesis is on extracting physical models from the image analysis and synthesis co-evolution from the Ecole Nationale Supérieure de Techniques Avancées de Paris, France. He is currently a Senior Researcher with ARMINES. He is also a Researcher with the Robotics Laboratory, with a special focusing on the image analysis for object and pattern recognition. He is also the Coordinator of a number of national and industrial research projects and he will be the person in charge for research and technological aspects for ARMINES for HELICoiD. His research interests include computer vision and pattern recognition, machine learning for image classification, and stochastic algorithms.

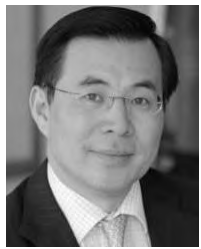

**GUANG-ZHONG YANG** (F'11) received the Ph.D. degree in computer science from Imperial College London, in 1991. He is currently the Director and the Co-Founder of The Hamlyn Centre for Robotic Surgery. He is also the Chairman of the UK-RAS Network. The mission of the UK-RAS Network is to provide academic leadership in RAS, expand collaboration with industry and integrate and coordinate activities of the EPSRC funded RAS capital facilities, Centres for Doctoral Training and partner universities across the UK. His current research interests include medical imaging, sensing, and robotics.

Prof. Yang is a Fellow of the Royal Academy of Engineering, IET, AIMBE, IAMBE, MICCAI, and CGI. He was a recipient of the Royal Society Research Merit Award and listed in The Times Eureka Top 100 in British Science. He was also a recipient of the CBE in the Queen's 2017 New Year Honour for his contribution to biomedical engineering.

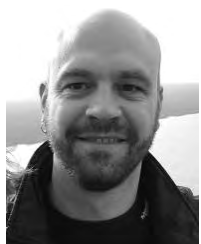

**RUBÉN SALVADOR** received the M.Sc. degree in 2007 and the Ph.D. degree in industrial electronics from the Universidad Politécnica de Madrid (UPM), in 2015, where he is currently an Assistant Professor with the Department of Telematics and Electronics Engineering and a Researcher Associate with the Center on Software Technologies and Multimedia Systems for Sustainability (CITSEM-UPM).

He was a Research Assistant with the Center of Industrial Electronics, UPM, from 2006 to 2011, and also with the Intelligent Vehicle Systems Division, University Institute for Automobile Research (INSIA), UPM, from 2005 to 2006. In 2009, he was a Visiting Research Student for 4 months with the Department of Computer Systems, Brno University of Technology. In 2017, he was a Visiting Professor for 5 months with IETR/INSA, Rennes. He has authored or co-authored around 40 peer reviewed publications in international journals/conferences and one book chapter. He has participated in nine EU/national research projects and nine industrial projects. His research interests include embedded, reconfigurable, and heterogeneous systems for parallel computing in FPGAs and manycore accelerators, system self-adaptation, and evolvable hardware. He focuses in design methodologies and architectures for highly parallel accelerators in embedded and high performance computing systems in the biomedical and space fields. The application domain spans hyperspectral image processing, machine learning, and evolutionary computing techniques for diagnostic imaging and self-adaptation in harsh environments.

He serves as a TPC member for various international conferences. He acts as a Reviewer in a number of international journals/conferences.

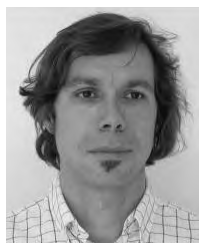

**EDUARDO JUÁREZ** (M'96) received the Ingeniero de Telecomunicación degree from the Universidad Politécnica de Madrid (UPM), Madrid, Spain, in 1993, and the Docteur ès Sciences Techniques degree from the École Polytechnique Fédéral de Lausanne (EPFL), Switzerland, in 2003. In 1994, he joined the Digital Architecture Group (GAD), UPM, as a Researcher. In 1998, he joined the Integrated Systems Laboratory (LSI), EPFL, as an Assistant. In 2000, he joined Transwitch Corp., Switzerland, as a Senior System Engineer. In 2004, he joined the Electronic and Microelectronic Design Group (GDEM) as a Postdoctoral Researcher. Since 2013, he has been a Researcher with the Research Centre of Software Technologies and Multimedia Systems (CITSEM). His current interests include low power hyperspectral embedded imaging systems for health applications.

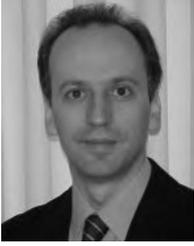

**CÉSAR SANZ** (S'87–M'88–SM'13) received the Ingeniero de Telecomunicación degree and the Doctor Ingeniero de Telecomunicación degree from the Universidad Politécnica de Madrid, Spain, in 1989 and 1998, respectively. Since 2008, he has been the Director of the E.U.I.T. de Telecomunicación, where he currently leads the Electronic and Microelectronic Design Group (GDEM) involved in R&D projects with Spanish and European companies and public institutions. His interests include microelectronic design applied to image coding, digital TV, digital video broadcasting, and hyperspectral imaging.

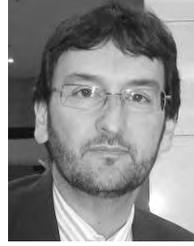

**ROBERTO SARMIENTO** was the Dean of the Faculty with the Telecommunication Engineering School, from 1994 to 1995 and a Vice Chancellor for Academic Affairs and a Staff with ULPGC, from 1998 to 2003. In 1993, he was a Visiting Professor with The University of Adelaida, South Australia, and later at the University of Edith Cowan, Australia. He is currently a Full Professor with the Telecommunication Engineering School, University of Las Palmas de Gran Canaria, Spain, in electronic engineering. He is also a Founder of the Research Institute for Applied Microelectronics (IUMA), where he is also the Director of the Integrated Systems Design Division. Since 1990, he has been publishing more than 50 journal papers and book chapters and more than 140 conference papers. He has been awarded with four six years research periods by the National Agency for the Research Activity Evaluation, Spain. He has participated in more than 45 projects and research programmes funded by public and private organizations, from which he has been leader researcher in 16 of them. Between these projects, it has special mention those funded by the European Union like GARDEN and the GRASS workgroup and the funded by the European Spatial Agency TRPAO8032. He has gotten several agreements with companies for the design of high performance integrated circuits, where the most important are those performed with Vitesse Semiconductor Corporation, California, Ensilica Ltd., U.K., and Thales Alenia Space, Spain.

● ● ●